\documentclass[letterpaper]{jpconf}
\usepackage{graphicx}
\begin{document}
\title{A realist interpretation of quantum mechanics based on
undecidability due to gravity}

\author{Rodolfo Gambini$^1$, Luis Pedro Garc\'{\i}a-Pintos$^1$, Jorge
  Pullin$^2$} 

\address{$^1$\,Instituto de F\'{\i}sica, Facultad de Ciencias,
  Universidad
  de la Rep\'ublica, Igu\'a 4225, CP 11400 Montevideo, Uruguay\\
  $^2$\, Department of Physics and Astronomy, Louisiana State
  University, Baton Rouge, LA 70803-4001}

\date{October 16th 2010}

\begin{abstract}
We summarize several recent developments suggesting that solving the
problem of time in quantum gravity leads to a solution of the 
measurement problem in quantum mechanics. This approach has been
informally called ``the Montevideo interpretation''. In particular
we discuss why definitions in this approach are not {\em for all
  practical purposes} (fapp) and how the problem of outcomes is resolved.
\end{abstract}

\section{Introduction}

The problem of measurement in quantum mechanics arises in standard
treatments as the requirement of a reduction process when a
measurement takes place.  Such process is not contained within the
unitary evolution of the quantum theory but has to be postulated
externally and is not unitary.  It is usually justified through the
interaction with a large, classical measuring device.

Two main objections have been levied onto two aspects of the solution
of the problem of measurement through decoherence.

1) Since the evolution of the system plus environment is unitary, the
coherence could potentially be regained.  

2) The second criticism has to do with the fact that in a picture
where evolution is unitary ``nothing ever occurs''. This is Bell's
``and/or'' problem. The final reduced density matrix of the system
plus the measurement device will describe a set of coexistent options
and not alternative options with definite probabilities.  

We shall argue that a solution to the problem of time in quantum
gravity leads to a modification of quantum mechanics such that the two
above objections can be overcome. The result is an objective
description of quantum mechanics that is compatible with unitarity and
does not require a reduction postulate.

The plan of the article is as follows. In the next section we briefly
describe the proposed solution to the problem of time in quantum
gravity. In section III we discuss how such solution leads to a
modification of Schr\"odinger's equation. In section IV we discuss the
solution to the objections to environmental decoherence. In section V
we argue that fundamental undecidability leads to the solution of the
problem of outcomes and macro-objectification. We end with a discussion.

\section{The problem of time in quantum gravity}

Quantum gravity is expected to be most relevant in situations where
one cannot assume that one has a clock external to the system under
study (for instance when one considers the universe as a whole as is
done in quantum cosmology). As a consequence of that one has to choose
as clock one of the variables of the system under study. Since the 
variables of the system under study are all represented by quantum
operators, time is not going to be a classical parameter as in
ordinary quantum mechanics.

In fact one can choose to describe ordinary quantum mechanics in terms
of a ``real clock'' i.e. a variable represented by a quantum operator
subject to quantum fluctuations, etc. There usually is an objection
that no monotonous variable can be conjugate to the Hamiltonian and
this is seen as an obstruction to time being represented by a physical
variable. But this is easily solved: one picks variables that work
well as clocks for a while. If one needs a longer lasting clock one
adds another variable when the first stopped being monotonous pretty
much like the hour hand on a clock does when the minute hand goes
around a complete circle. The way to describe quantum mechanics 
with a real clock is relationally. You pick an observable, let us call
it $T(t)$ and then pick a set of observable quantities we want to study that
commute with $T(t)$, i.e. $O^1(t),\ldots,O^N(t)$. Here we denote by
$t$ the ideal classical time that appears in the ordinary
Schr\"odinger equation. One then computes the conditional probability,
\begin{equation}
  P\left(O^i=O^i_0\vert T=T_0\right) = 
\lim_{\tau\to\infty} 
{\int_{-\tau}^{\tau} dt
{\rm Tr}\left( P_{O^i_0}(t) P_{T_0}(t) \rho P_{T_0}(t)\right) \over
\int_{-\tau}^{\tau} dt
{\rm Tr}\left(  P_{T_0}(t) \rho \right)},\nonumber
\end{equation}
where $P_{O^i_0}(t)$ is the projector on the eigenspace
associated with the eigenvalue $Q^i_0$ at time $t$ and
similarly for $P_{T_0}(t)$. These conditional probabilities are positive
and add to one. If one chooses operators with continuum spectrum such
probability of giving a definite value vanishes, one has to ask about
the probability of being in an interval around the eigenvalue and the
above expression is unchanged. 

There is a technical problem in carrying out the above construction in
theories like general relativity that are generally covariant and it
is the construction of observables to use in the construction. This
was for many years a stumbling block for carrying out this
program. Recently we proposed to use evolving constants of the motion
as the observables and this construction has appeared satisfactory in
model systems that could not be treated before. Since we are orienting
this article towards readers not primarily interested in quantum
gravity we omit the details here and refer to our paper on the subject
\cite{time}.

\section{A modified Schr\"odinger equation}

An important question is how the conditional probabilities we
introduced in the previous section evolve. We have been able to show
that if one make some judicious assumptions, namely, that the clock
does not interact with the system, that the clock is in a highly
classical state (a coherent state where the ``hand'' of the clock is
sharply peaked in space and moves in a monotonous way), then one can
define a density matrix labeled by the eigenvalues of $T$, 
\begin{equation}
\rho(T) \equiv \int_{-\infty}^\infty U_{\rm sys}(t) \rho_{\rm sys} 
 U_{\rm sys}(t)^\dagger {\cal P}_t(T)
\end{equation}
where the ``sys'' subscript indicates that it is the density matrix
and the evolution operator corresponding to the system under study, it
does not include the clock. We are assuming the density matrix of the
whole universe is of the form, $\rho=\rho_{\rm sys}\otimes \rho_{\rm
  clock}$. The probability ${\cal P}_t(T)$ is the probability that 
the clock reads $T$ when the ideal Schr\"odinger time is $t$. Such
probabilities are not directly observable, that is the reason the
variable $t$ always appears integrated. The explicit expression for
the probability is,
\begin{equation}
  {\cal P}_t(T) = \frac{{\rm Tr}\left(P_T(t) \rho\right)}
{\int_{-\infty}^\infty dt' {\rm Tr}\left(P_T(t')\rho\right)}.
\end{equation}

If one assumes one has a clock that follows the ideal Schr\"odinger
time perfectly, then ${\cal P}_t(T)=\delta(t-T)$. In reality there
will be departures. If we assume the departures are very small,
\begin{equation}
{\cal P}_t(T)=\delta(t-T)
=\delta(T-t)+ a(T)\delta'(T-t)+b(T)\delta''(T-t)+\ldots,
\end{equation}
one can show that the density matrix satisfies an approximate 
Schr\"odinger equation (for a pedagogical discussion of the
derivation, see \cite{obregon}),
\begin{equation}
{\partial \rho(T)\over \partial T} =i [\rho(T),H] +\sigma(T) [H,[H,\rho(T)]].
\end{equation}
where $\sigma(T)=\partial b(T)/\partial T$. This type of equation is a
particular case of the Lindblad equation considered in decoherence,
but in this particular form it has the virtue that it conserves
energy. The effect of the extra term is to make the off diagonal
elements of the density matrix in the energy basis go to zero
exponentially. That is, the evolution in terms of the real clock is
not unitary. The underlying evolution in terms of the idealized time
$t$ is, but the ``real clock'' $T$ cannot keep track of it accurately
enough to keep a unitary evolution. After some time for a given $T$
one will have a distribution of possible $t$'s associate with it and
even if one started with a pure state one ends with a superposition of
pure states for a given value of $T$. In what follows each time we
refer to unitarity we will be speaking of the evolution in the
underlying unobservable parameter. The observable evolution depends on
the particular implementation of the clock we are taking and it is not
unitary.

At this point it is worthwhile asking: can one make the effect
arbitrarily small by choosing better clocks? That requires to estimate
the minimum uncertainty one can have in a clock. A full discussion
would require more understanding of quantum gravity that we have right
now. There have been several heuristic estimates in the literature
\cite{ng} and
some controversy surrounding them. We prefer to just consider
generically that the uncertainty in the measurement of time has to
grow with the observed time and if the effect is due to gravity,
Planck's time $t_{\rm Planck}=10^{-44}s$ should be involved. This
leads to postulate an uncertainty in the measurement of a period of
time $T$ of the form,
\begin{equation}
  \delta T = t_{\rm Planck}^{1-a} T^{a}.
\end{equation}
The heuristic arguments we mentioned above yield $a=1/3$ or $a=1/2$
but in practice this makes little difference. As long as the clock
error increases with the time measured, there will be a fundamental loss
of coherence due to the use of real clocks.

A similar effect is introduced by the use of ``real rods'' to measure
space. Such a discussion requires quantum field theory and had not
been developed fully so we refer the reader to reference \cite{spatial}.

\section{Solution to the objections to environmental decoherence}

The presence of a fundamental source of loss of coherence helps deal
with the objections levied against the solution of the problem of
measurement through environmental decoherence. The two main objections
can be characterized as ``the information is still there'' and ``the and/or''
problem or ``problem of outcomes''. 
We discuss them in the following two subsections.

\subsection{Impossibility of recovering the information}

This objection is as follows: although a quantum system interacting
with an environment with many degrees of freedom will very likely give
the appearance that the initial quantum coherence of the system is
lost, the information about the original superposition could be
recovered, for instance, by carrying out a measurement that includes
the environment.  The fact that such measurements are hard to carry
out in practice does not prevent the issue from existing as a
conceptual problem. 

To analyze this point in some detail, as is customary in discussions
of decoherence, let us consider a variation of a model proposed by
Zurek \cite{zurek} in which the quantum system, the environment and
the measuring apparatus are under control. The system consists of a
cavity with a uniform magnetic field in the $z$ (vertical)
direction. Inside the cavity is a spin $S$ that represents the
``needle'' of the measuring device. A steady flux of spins is pumped
into the cavity in a horizontal stream and each spin interacts with $S$
a finite amount of time $\tau$. The stream constitutes ``the
environment''. We also assume the spins of the environment are
separated enough to avoid considering interactions among them. The
interaction Hamiltonian is given by,
\begin{equation}
\hat{H}^{\rm int}_k = f_k \left(
\hat{S}_x \hat{S}_x^k +
\hat{S}_y \hat{S}_y^k +
\hat{S}_z \hat{S}_z^k\right),
\end{equation}
with $f_k$ the coupling constants. 
The Hamiltonian due to the presence of the magnetic field when the $k$-th particle is in
the cavity is,
\begin{equation}
\hat{H}^B_k = \gamma_1 B \hat{S}_z\otimes \hat{I} _k +\gamma_2
B\hat{I}\otimes S^{k}_z,
\end{equation}
where $\hat{I}$ is the identity matrix acting on the Hilbert space of
the needle and $I_k$ is the identity in the Hilbert space of the
$k$-th particle. The introduction of a constant magnetic field
pointing in a given direction, is in order to have a definite pointer
basis.

We have discussed this model in detail in \cite{foundations2} so we
will not repeat all details here. We outline the main
result. d'Espagnat \cite{despagnat} 
suggested that for systems like this one can define
a global observable,
\begin{equation}
\hat{M} \equiv \hat{S}_x \otimes\prod_k^N \hat{S}^k_x.
\end{equation}
This observable has the property that if one starts from a normalized
initial state given by,
\begin{equation}
\vert\Psi(0)\rangle = \left(a\vert+\rangle + b\vert-\rangle\right) \prod_{k=1}^N \otimes \left[
\alpha_k\vert+\rangle_k +\beta_k \vert-\rangle_k \right],
\end{equation}
where $\vert\pm\rangle$ are eigenstates of $\sigma_z$, and
computes the expectation value of the observable 
in a state that has evolved unitarily one gets,
\begin{equation}
\langle \psi\vert M\vert\psi\rangle =
ab^*\prod_k^N\left[\alpha_k\beta_k^* +\alpha^*_k\beta_k\right]
e^{-2i\Omega_k\tau} + a^*b
\prod_k^N\left[\alpha_k\beta_k^* +\alpha^*_k\beta_k\right]
e^{2i\Omega_k\tau},
\end{equation}
where $ \Omega_k = \sqrt { 4 f_k^2 + B^2 (\gamma_1 - \gamma_2) ^2 }$
with $\gamma_1$ the magnetic moment of the spins of the environment
and $\gamma_2$ that of the ``needle''.  On the other hand, if one
assumes a collapse of the wavefunction has occurred one gets that
\begin{equation}
  \langle \psi\vert M\vert\psi\rangle =0.
\end{equation}
If one now studies the same observable but assuming the modified
Schr\"odinger evolution we discussed in the previous section one gets, 
\begin{eqnarray}
\langle \hat{M}\rangle &=&
a b^* e^{-i2 N \Omega T} e^{-4 N B^2 (\gamma_1-\gamma_2)^2 \theta}
\prod_{k}^N \left[ \alpha_k \beta_k^* e^{-16 B^2\gamma_1\gamma_2\theta}
+\alpha_k^* \beta_k\right]\\
&&+b a^*
e^{i2 N \Omega T} e^{-4 N B^2 (\gamma_1-\gamma_2)^2 \theta}
\prod_{k}^N \left[ \alpha_k \beta_k^*
+\alpha_k^* \beta_ke^{-16 B^2\gamma_1\gamma_2\theta}\right]
\end{eqnarray}
where $\Omega= B(\gamma_1 -\gamma_2)$.  
We see the expectation value
is exponentially damped with an exponent that is
large. Actually putting in realistic numbers one concludes that if one
has an ``environment'' with more than $10^7$ particles the expectation
value decays too fast to be measured within the accuracies of the
experiment, as we will discuss in the next subsection.

Due to the fundamental decoherence induced by quantum clocks the
expectation value of the observables exponentially decreases and is
more and more difficult to distinguish from the vanishing value
resulting from collapse.  A similar analysis allows to show that
``revivals'' \cite{foundations} of coherence of the wavefunction of
the system plus the measurement device (but excluding the environment)
are also prevented by the modified evolution.  When the multi-periodic
functions in the coherences tend to take again the original value
after a Poincar\'e time of recurrence the exponential decay for
sufficiently large systems completely hides the revival under the
noise amplitude.

Thus, the difficulties found in testing macroscopic superpositions in
a measurement process are enhanced by the corrections resulting of the
use of physical clocks.

\subsection{Fundamental undecidability and the problem of outcomes}

To address the second objection to the decoherence solution to the
measurement problem, the ``and-or'' problem or as others put it
``nothing ever happens'' we will introduce a criterion to determine
when an event takes place in a theory where evolution is always
unitary. 

When one takes into account the way that time enters in general
covariant systems including the quantum fluctuations of the clock, the
unitary evolution of the total system (system plus apparatus plus
environment) becomes indistinguishable from a statistical mixture of
all the alternative states that correspond to each of the possible
outcomes of a measuring device. By indistinguishable we mean that all
physical predictions of both states cannot be distinguished due to
fundamental limitations in measurement.  We call such situation
``undecidability''. We will claim that an event takes places when
undecidability has been reached.  Therefore undecidability is truly a
comparison between the quasi-unitary evolution in physical time
described by the Lindblad equation and the statistical mixture of the
states associated with the different outcomes of the measuring device.

The fact that in the definition of undecidability we had to invoke
fundamental limitations in measurement raises the question is the
definition is ``for all practical purposes'' only, or if it is
fundamental in nature. That is, it is not due to a technological
limitation, but due to limitations imposed by the laws of physics. We
are going to show that undecidability is not only for all practical
purposes (FAPP) but fundamental (see \cite{fapp} for details).  From
the previous discussion one can gather that as one considers
environments with a larger number of degrees of freedom and as longer
time measurements are considered, distinguishing between collapse and
unitary evolution becomes harder. But is this enough to be a
fundamental claim?

Given the expectation value we discussed 
\begin{equation}
  <\hat{M}> \sim \exp\left(-6 N B^2(\gamma_1-\gamma_2)^2 T_{\rm
      Planck}^{4/3} 
\tau^{2/3}\right)
\end{equation}
is it possible for a big enough ensemble to distinguish it from zero?

Brukner and Kofler \cite{brko} have recently proved that from a very
general quantum mechanical analysis together with bounds from special
and general relativity there is a fundamental uncertainty in the
measurements of angles associated with quantum mechanical spin, 
even if one uses a measuring device of the size
of the observable Universe. The bound is given by $\Delta\theta\ge
\ell_{\rm Planck}/R$ with $\ell_{\rm Planck}$ the Planck length and
$R$ the radius of the universe. This uncertainty in the measurement of
angles in our example implies there is an uncertainty in the
measurement of $<M>$. If that uncertainty is bigger than the value of
$<M>$ one can measure there is no way of distinguishing a collapse
from a unitary evolution even in principle. For the model in question
this occurs for $N\sim 10^7$ spins. So one sees that even if one 
considers ensembles of growing size, this does not help in
distinguishing the expectation value in the case of collapse or
unitary evolution since there are fundamental limitations to
measurement. It is like attempting to measure a distance with an
accuracy greater than the smallest spacing in a ruler. No matter how
many measurements one makes, the error will be at least as large as
the smallest spacing on the ruler.

The problem of macro-objectification of properties may be described
according with Ghirardi \cite{Ghirardi} as follows: ``how, when, and
under what conditions do definite macroscopic properties emerge (in
accordance with our daily experience) for systems that, when all is
said and done, we have no good reasons for thinking they are
fundamentally different from the micro-systems of which they are
composed?''  We think that undecidability provides an answer to this
problem.  If a system suffers an interaction with an environment such
that the density matrix is such that all the physical predictions of
such a state cannot be distinguished from those of a statistical
mixture of the states corresponding to different outcomes of the
collapsed system, we will claim that an event took place.  A
fundamental ingredient is that in order to distinguish the density
matrices one will be limited by fundamental limitations on the
measurements of physical quantities and preparations of physical
states. A detailed implementation will depend on the system under
study. An example is the limitation on the measurements of spins of
Brukner and Kofler we discussed in this section. Notice that for a
quantum micro-system in isolation, events would not occur. However for
a quantum system interacting with an environment, events will be
plentiful. The same goes for a system being measured by a macroscopic
measuring device.

\section{Discussion}

Gravity fundamentally limits how accurate our measurements of space
and time can be.  This requires reformulating quantum mechanics in
terms of real clocks and rods, that have errors in their
measurements. The resulting picture of quantum theory is one where
there is a fundamental loss of quantum coherence: pure states evolve
into mixed states as our clocks and rods cannot keep track of the
unitary evolution.  This eliminates the problem of revivals and global
observables in the solution to the measurement problem through
environmental decoherence: just waiting longer does not improve things
as more quantum coherence is lost.  One also has a definition for when
an event takes place: when the fundamental loss of coherence is such
that one cannot distinguish the unitary evolution from a reduction (in
the sense previously explained), an
event has taken place.  We call this situation ``undecidability'' between
reduction and unitary evolution. One therefore has a complete,
objective formulation of quantum mechanics with a unitary evolution
and a notion of event that does not imply the collapse of the
wavefunction. We call it the Montevideo interpretation of quantum
mechanics. There even exists an axiomatic formulation of the
theory \cite{axiomatic}. We refer the reader to the references for
further details.

\ack Part of this work was done in collaboration with Rafael Porto and
Sebasti\'an Torterolo.  This work was supported in part by grant
NSF-PHY-0650715, funds of the Hearne Institute for Theoretical
Physics, CCT-LSU, Pedeciba and ANII PDT63/076.


\section*{References:}
\begin{thebibliography}{9}
\bibitem{time}
  R.~Gambini, R.~A.~Porto, J.~Pullin and S.~Torterolo,
  Phys.\ Rev.\  D {\bf 79}, 041501 (2009)
  [arXiv:0809.4235 [gr-qc]].

\bibitem{obregon}
  R.~Gambini, R.~Porto and J.~Pullin,
  Gen.\ Rel.\ Grav.\  {\bf 39}, 1143 (2007) [arXiv:gr-qc/0603090].

\bibitem{ng}
F. K\'arolhy\'azy, A. Frenkel, B. Luk\'acs in ``Quantum concepts in
space and time'' R. Penrose and C. Isham, editors, Oxford University
Press, Oxford (1986); Y.~J.~Ng and H.~van Dam, Annals N.\ Y.\ Acad.\
Sci.\ {\bf 755}, 579 (1995) [arXiv:hep-th/9406110]; Mod.\ Phys.\
Lett.\ A {\bf 9}, 335 (1994); G.~Amelino-Camelia, 
Measurability Of Space-Time Distances In The Semiclassical
3415 (1994) [arXiv:gr-qc/9603014]; S. Lloyd, J. Ng, Scientific
American, November (2004).

\bibitem{spatial}
  R.~Gambini, R.~A.~Porto, J.~Pullin,
  Phys.\ Lett.\  {\bf A372}, 1213-1218 (2008).
  [arXiv:0708.2935 [quant-ph]]

\bibitem{zurek} W. Zurek, Phys. Rev. {\bf D26}, 1862 (1982).

\bibitem{foundations2} 
  R.~Gambini, L.~P.~G.~Pintos, J.~Pullin,
  Found.\ Phys.\  {\bf 40}, 93-115 (2010).
  [arXiv:0905.4222 [quant-ph]].

\bibitem{despagnat}
B. d'Espagnat ``Veiled reality'', Addison Wesley,
  New York (1995).

\bibitem{foundations}
  R.~Gambini, J.~Pullin,
  Found.\ Phys.\  {\bf 37}, 1074-1092 (2007).
  [quant-ph/0608243].


\bibitem{fapp}
 R.~Gambini, L.~P.~Garcia-Pintos, J.~Pullin,
    [arXiv:1009.3817 [quant-ph]].

\bibitem{brko}
C. Brukner, J. Kofler, ``Are there fundamental
  limits for observing quantum phenomena from within quantum
  theory?'', arXiv:1009.2654 [quan-ph]

\bibitem{Ghirardi} 
G. Ghirardi, ``Sneaking a look at God's cards'' Princeton University
Press, Princeton, NJ (2007).

\bibitem{axiomatic}
  R.~Gambini, L.~P.~Garcia-Pintos, J.~Pullin,
  ``Complete quantum mechanics: An Axiomatic formulation of the Montevideo interpretation,''  
  [arXiv:1002.4209 [quant-ph]].
\end{thebibliography}
\end{document}